\documentclass[letterpaper, preprint, paper,11pt]{AAS}	

\usepackage{bm}
\usepackage{amsmath}
\usepackage[colorlinks=true, pdfstartview=FitV, linkcolor=black, citecolor= black, urlcolor= black]{hyperref}
\usepackage{overcite}
\usepackage{footnpag}			      	

\usepackage{array}
\usepackage{subcaption}
\usepackage{multirow}
\usepackage{amssymb}
\usepackage[normalem]{ulem}
\usepackage{soul}
\usepackage{cancel}
\usepackage{xcolor}

\PaperNumber{25-727}

\begin{document}

\title{Fuzzy-Based Control Method for Autonomous Spacecraft Inspection with Minimal Fuel Consumption}

\author{Daegyun Choi\thanks{Postdoctoral Researcher, Department of Aerospace Engineering \& Engineering Mechanics, University of Cincinnati, Cincinnati, OH 45221, USA.},  
Donghoon Kim\thanks{Assistant Professor, Department of Aerospace Engineering \& Engineering Mechanics, University of Cincinnati, Cincinnati, OH 45221, USA.}, \ and 
Henzeh Leeghim\thanks{Professor, Department of Aerospace Engineering, Chosun University, Gwangju 61452, Republic of Korea.}
}

\maketitle{}

\begin{abstract}
This study explores an energy-efficient control strategy for spacecraft inspection using a fuzzy inference system combined with a bio-inspired optimization technique to incorporate learning capability into the control process. The optimized fuzzy controller produces a minimally fuel-consuming force while maintaining reliable inspection within constraints, such as illumination, restricted field of view, thrust limits, and safe regions. The performance of the proposed control strategy is validated through Monte Carlo simulations.
\end{abstract}

\section{Introduction}
As the number of spacecraft in orbit grows and ages, malfunctions due to harsh space environments or component failures are becoming more frequent. 
This trend has led to a significant increase in interest in on-orbit servicing (OOS). The market for OOS is rapidly expanding, projected to grow from \$2.4 billion in 2023 to \$5.1 billion by 2030, at an 11.5\% compound annual growth rate \cite{oos}. This growth is largely driven by the rising need to extend spacecraft lifespans and increased investment in the global OOS market. A notable example is Northrop Grumman's Mission Extension Vehicle-1 (MEV-1), which took over control of the nearly fuel-depleted Intelsat-901 communication satellite \cite{MEV}.

When a client spacecraft's status is known, servicing missions can proceed without an initial inspection. However, if this information is unavailable, the client's health and status must be assessed before any servicing is provided. Beyond immediate servicing needs, periodic monitoring of spacecraft is crucial for early detection of potential issues like damage or component malfunctions. This proactive approach helps prevent mission failures and enables informed decisions about repairs or maintenance, ultimately extending spacecraft operational life. Such monitoring can be achieved by a smaller spacecraft (deputy) inspecting a larger spacecraft (chief). An example of this is the European Space Agency's Space Rider system \cite{spaceRider}, currently under development, which features an uncrewed robotic laboratory with a small satellite designed for inspecting the main spacecraft. Therefore, autonomous spacecraft inspection is vital for preventing problems in space, advancing OOS missions, and extending spacecraft mission life.

Autonomous spacecraft inspection presents significant challenges due to numerous constraints, making it a complex problem. These challenges include coordinating the six degrees of freedom for attitude and position control of the spacecraft, the deputy's limited capabilities (e.g., actuator limits, restricted sensor field of view (FOV)), avoiding collisions while ensuring reliable inspection, and managing illumination. Researchers have explored these problems, often using optimal control approaches, such as a combination of 
traditional multistart methods with sequential quadratic programming \cite{Laguardia2024_toc} or model predictive control \cite{Nakka2021_mpc, Fu2022_mpc}, aiming to minimize energy consumption while considering various constraints. However, as the complexity of the problem increases, these optimal control methods demand substantial computational power, making real-time implementation difficult.

Consequently, many researchers are developing artificial intelligence (AI)-based control techniques to address the multi-constrained inspection problem. Reinforcement learning (RL) is a widely adopted AI approach for spacecraft inspection \cite{Wijk2024,Hibbard2023_rl,Dunlap2024_rl}. It can handle complex problems with multiple constraints and provide real-time solutions \cite{Yoo2021_rl}, though it requires extensive training. However, a significant drawback of RL is its lack of interpretability in decision-making, which makes it challenging for control engineers to understand why specific actions are taken, especially if they lead to undesirable outcomes.

To overcome the interpretability limitations of RL, this work proposes a fuzzy inference system (FIS)-driven control approach. The FIS offers a transparent decision-making process through 
rules based on 
If-Then statements, mapping inputs to outputs. It can also manage complex problems with low computational cost. However, designing the internal structure of FISs to achieve a near-optimal solution is difficult. To address this, this work integrates an optimization technique into the FIS to automatically determine its internal parameters. In this research, the FISs will determine the control force for the deputy spacecraft, aiming to minimize fuel consumption during chief spacecraft inspection while adhering to the aforementioned constraints.


\section{Problem Formulation}
This study addresses autonomous spacecraft inspection, in which a deputy spacecraft inspects a chief spacecraft with minimal energy consumption (herein referred to as delta-v consumption), while considering the deputy's limited capabilities, mission success, and environmental factors. 

To determine the deputy's optimized control input for successful mission execution, the objective function is defined as
\begin{equation}\label{eq:obj}
    \mathcal{L} = \Delta v, 
\end{equation}
subject to multiple equality and inequality constraints, including the deputy's dynamics and kinematics, inspection rate, collision avoidance, thrust limits, restricted FOV, and illumination conditions. 

Note that $\Delta v$ is the fuel consumption of the deputy, defined as
\begin{equation}
    \Delta v= \sum_{k=1}^{N}    \frac{|f_x(k)|+|f_y(k)|+|f_z(k)|}{m_d} T,
\end{equation}
where $T$ is the time interval, $N$ is the total number of data, $f_i(k)$ for $i=x$, $y$, $z$ is the control force along each axis at time step $k$, and $m_d$ is the deputy's mass. 

The details of the constraints are explained in the following sections.

\subsection{Dynamics and Kinematics}
\begin{figure}
    \centering
    \includegraphics[width=0.7\linewidth]{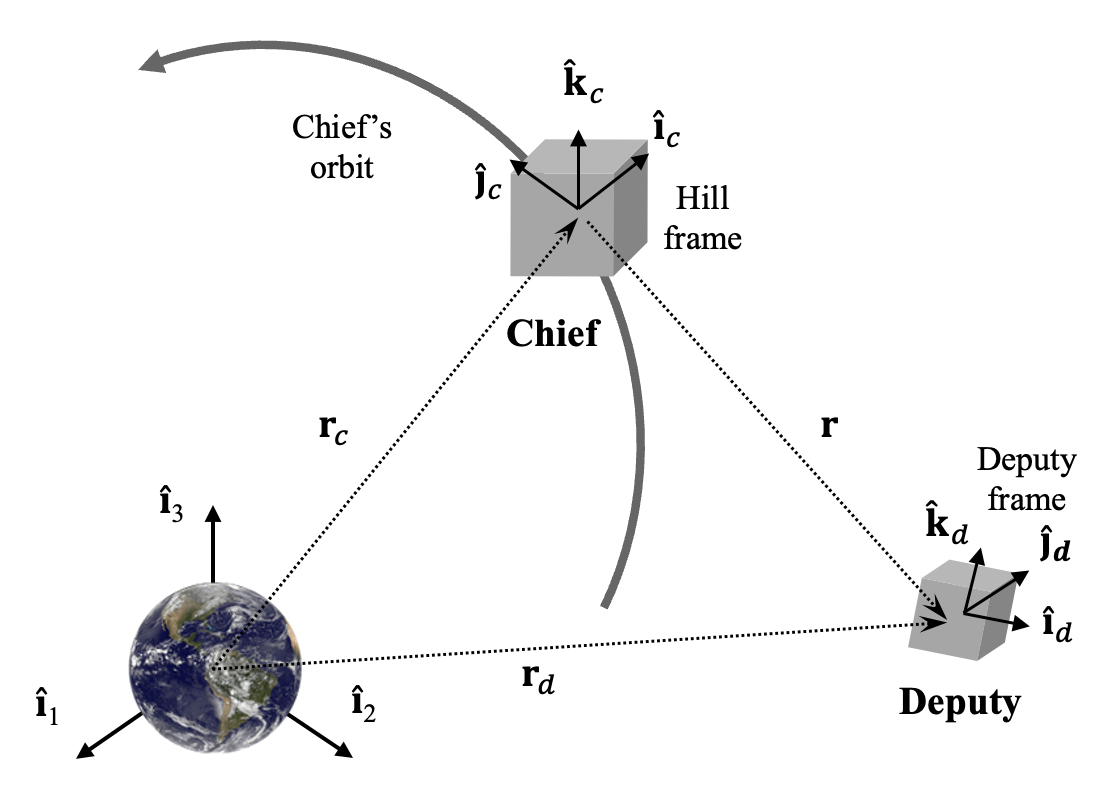}
    \caption{Definition of the Reference Frames and Position Vectors for Each Spacecraft}
    \label{fig:frame}
\end{figure}

To describe the motion of the chief and deputy, the Earth-centered inertial frame, as well as body-fixed frames attached to both the chief and deputy, are defined as shown in Figure~\ref{fig:frame}. 
The Hill frame is first defined, originating at the chief's center of mass. The unit vector $\hat{\bf i}_c$ is aligned with the position vector from the Earth's center to the chief's center of mass. The unit vector $\hat{\bf k}_c$ points in the direction normal to the chief's orbital plane, and $\hat{\bf j}_c$ is determined by the right-handed triad. In addition, the deputy frame is initially aligned with the Hill frame, and its origin is located at the deputy's center of mass. 

While the position vectors of the chief and deputy with respect to the Earth's center are defined as ${\bf r}_c\in\mathbb{R}^{3}$ and ${\bf r}_d\in\mathbb{R}^{3}$, the relative position vector of the deputy with respect to the chief is defined as ${\bf r}=[x,\ y,\ z]^T\in\mathbb{R}^3$. Assuming that the chief is on a circular orbit and $||{\bf r}||/||{\bf r}_c|| \ll 1 $, the deputy's relative translational motion with respect to the chief, expressed in the Hill frame, is described
by \cite{Schaub2003analytical}
\begin{align}
    \ddot{x} - 3n^2x - 2n\dot{y} & = (f_x + d_x )/m_d,\\
    \ddot{y} + 2n\dot{x} & = (f_y + d_y )/m_d,\\
    \ddot{z} + n^2 z & = (f_z + d_z )/m_d,
\end{align}
where $m_d$ is the deputy's mass, $n$ is the chief's mean motion, $f_i$ and $d_i$ ($i=x$, $y$, and $z$) are the {deputy's} control 
and disturbance forces, respectively, in the Hill frame. 
The detailed information to determine the control force is described in a later section.

In this work, the deputy's relative attitude with respect to the chief is represented using quaternions, defined as
${\bf q} = [q_1,\ q_2,\ q_3,\ q_4]^T = [{\bf q}_v^T,\ q_4]^T\in\mathbb{R}^4$, with the holonomic constraint ${\bf q}^T{\bf q} = 1$. Using this quaternion representation, the direction cosine matrix $C_{DC}$, which maps the chief (or Hill) frame to the deputy frame, is given by \cite{Schaub2003analytical}
\begin{equation}
    C_{DC} = \begin{bmatrix}
        q_1^2-q_2^2-q_3^2+q_4^2 & 2(q_1 q_2 + q_3 q_4) & 2(q_1 q_3 - q_2 q_4) \\
        2(q_1 q_2 - q_3 q_4) & -q_1^2+q_2^2-q_3^2+q_4^2 & 2(q_2 q_3 + q_1 q_4) \\
        2(q_1 q_3 + q_2 q_4) & 2(q_2 q_3 - q_1 q_4) & q_1^2-q_2^2+q_3^2+q_4^2
    \end{bmatrix}.
\end{equation}
The kinematic differential equations of the quaternions are given by \cite{Wie1989}
\begin{align}
    \dot{{\bf q}}_v & = -\frac{1}{2}[\bm\omega^\times]{\bf q}_v + \frac{1}{2}q_4\bm\omega,\\
    \dot{q}_4 & = -\frac{1}{2}\bm\omega^T{\bf q}_v.  
\end{align}
Here, [$\bm\rho^\times$] is the skew-symmetric matrix for a generic vector $\bm\rho = [\rho_1,\ \rho_2,\ \rho_3 ]^T\in\mathbb{R}^3$, defined as:
\begin{equation}
    [\bm\rho^\times] = \begin{bmatrix} 0 & -\rho_3 & \rho_2 \\ 
    \rho_3 & 0 & -\rho_1 \\ 
    -\rho_2 & \rho_1 & 0 \end{bmatrix}.
\end{equation}

The deputy's relative rotational motion relative to the chief is expressed by \cite{Li2018}
\begin{equation}
    J_d \dot{\bm\omega} -A\bm\omega - {\bf h} = \bm\tau + \bm\tau_d,
\end{equation}
where
\begin{align}
A & = -J_d [(C_{DC}\bm\omega_c)^\times] - [(C_{DC}\bm\omega_c)^\times]J_d + [J_d(\bm\omega+C_{DC}\bm\omega_c)^\times],\\
{\bf h} & = -[(C_{DC}\bm\omega_c)^\times]J_d C_{DC}\bm\omega_c - J_d C_{DC}\dot{\bm\omega}_c.
\end{align}
Here, $\bm\omega = \bm\omega_d- C_{DC}\bm\omega_c\in\mathbb{R}^3$ is the relative angular velocity of the deputy with respect to the chief, $\bm\omega_d\in\mathbb{R}^3$ is the angular velocity of the deputy with respect to the inertial frame (expressed in the deputy frame), 
$\bm\omega_c\in\mathbb{R}^3$ is the angular velocity of the chief with respect to the inertial frame expressed in the Hill frame, 
$J_d\in\mathbb{R}^{3\times 3}$ is the inertia matrix of the deputy, $\bm\tau\in\mathbb{R}^{3}$ is the control torque of the deputy expressed in the deputy frame, and $\bm\tau_d\in\mathbb{R}^{3}$ is the disturbance torque acting on the deputy, respectively. 

In this work, to concentrate on analyzing translational motion control, the deputy's control input is determined by a PD controller, expressed as:
\begin{equation}
    \bm\tau = -K_p {\bf q}_e - K_d \bm\omega,
\end{equation}
where ${\bf q}_e\in\mathbb{R}^{3}$ is the quaternion error between the deputy and the chief, and $K_p\in\mathbb{R}^{3\times 3}$ and $K_d\in\mathbb{R}^{3\times 3}$ are the proportional and derivative gain matrices, respectively.

The chief's rotational motion relative to the inertial frame is defined by Euler's equation with no external torque, expressed as \cite{Schaub2003analytical}:
\begin{equation}
    J_c \dot{\bm\omega}_c + [\bm\omega_c^\times]J_c\bm\omega_c = {\bf 0},
\end{equation}
where $J_c\in\mathbb{R}^{3\times 3}$ is the inertia matrix of the chief.


\subsection{{Inspection Rate}}
\begin{figure}
    \centering
    \includegraphics[width=0.7\linewidth]{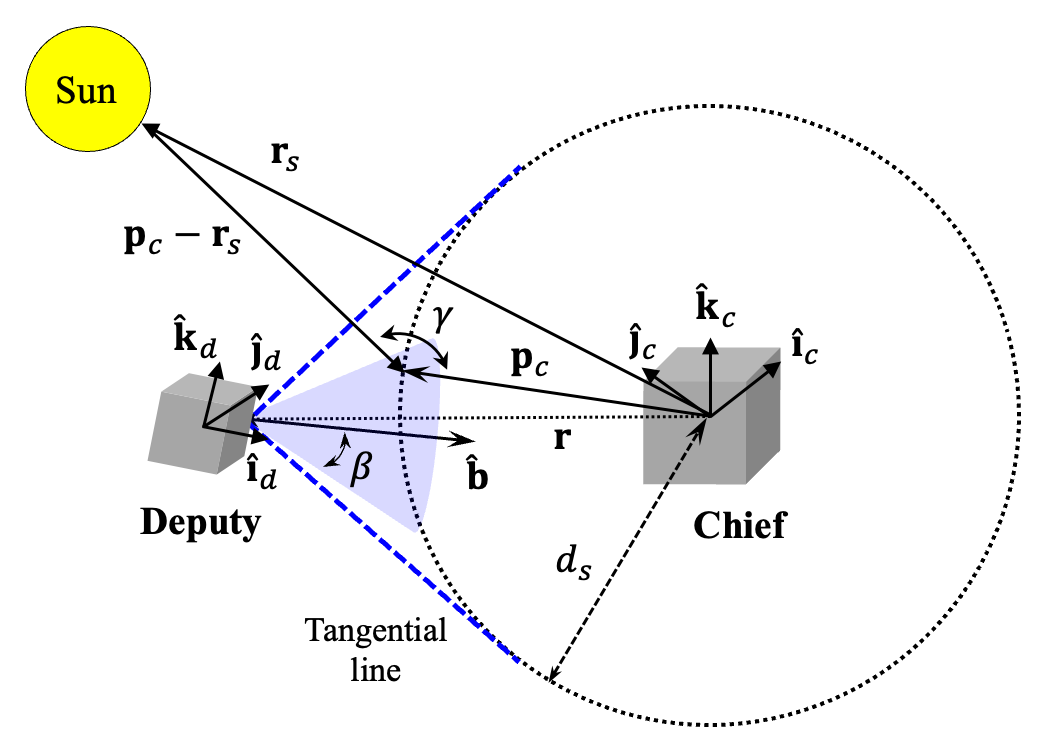}
    \caption{Schematic Diagram of the Deputy's FOV and Illumination}
    \label{fig:constraints}
\end{figure}

To generalize the inspection problem, this work defines inspection points around the chief (i.e., spherically distributed points at a distance of $d_s$ from the chief's center), as illustrated in Figure \ref{fig:constraints}. 
The success of the inspection of the chief can be quantified by evaluating the number of inspected points.
Based on the number of inspected points ($n_{\text{insp}}$) and the total number of inspectable points ($n_{\text{total}}$) {around the chief}, the inspection rate is {expressed }
as $\eta_{\text{insp}}=n_{\text{insp}}/n_{\text{total}}\times 100$. 
{Here, }$n_{\text{total}}$ is {determined by users. }
{In this work, }the mission is considered successful if the inspection rate {satisfies the following condition:}
\begin{equation}\label{eq:insp_rate}
    \eta_\text{threshold} \le \eta_\text{insp},
\end{equation}
{where $\eta_\text{threshold}$ is the inspection threshold.}

\subsection{{Collision Avoidance}}

For a successful and safe mission, potential collision risks between the two spacecraft must be avoided. To ensure the deputy's safe maneuvers, this work considers allowable relative distance ranges between the deputy and the chief. While the maximum relative distance is constrained to ensure reliable inspection, the minimum relative distance is defined to avoid potential collisions. Therefore, the deputy's position with respect to the chief must satisfy the following constraint:
\begin{equation}\label{eq:dist}
    d_\text{min} \le ||{\bf r}|| \le d_\text{max},
\end{equation}
where $d_\text{min}$ and $d_\text{max}$ are the minimum and maximum allowable relative distances, respectively, and $||{\bf r}||$ is the relative distance between the deputy and the chief.

\subsection{{Thrust Limits}}
The deputy's translational and rotational motions are controlled by actuation force and torque. Such actuations are typically constrained by the propulsion type and size of the thrusters for translational motion, and by the size and rotational speed of the momentum exchange devices for rotational motion. 

Therefore, this work employs thrust force and actuation torque limits during inspection to reflect realistic mission constraints, defined as:
\begin{align}
    |f_i| &\le f_\text{max},\\
    |\tau_i| & \le \tau_\text{max},
\end{align}
where $i=x$, $y$, and $z$, and 
$f_\text{max}$ and $\tau_\text{max}$ are the maximum thrust force {and torque} generated by the actuators.

\subsection{{Restricted FOV}}
This study considers visual inspection of the chief using optical sensors. Such sensors typically have a limited FOV, affecting inspection performance. To find points around the chief that are within the FOV, it must satisfy the following two conditions:
\begin{align} 
    \cos^{-1} \left(\frac{{\bf p}_c\cdot {\bf r}}{||{\bf p}_c|| \cdot ||{\bf r}||}\right) \le \cos^{-1} \left( \frac{d_c}{||{\bf r}||}\right), \label{eq:fov1} \\
    \cos \beta - \frac{({\bf p}_c-{\bf r})\cdot \hat{\bf b}}{||({\bf p}_c-{\bf r})\cdot \hat{\bf b}}\le0, \label{eq:fov2}
\end{align}
where $\hat{\bf b}$ is the boresight vector of the optical sensor installed on the deputy, and $\beta$ is the sensor's half-angle FOV. 
Equation \eqref{eq:fov1} ensures that the inspection point lies within the range generated by two tangential lines from the deputy to the surface of the spherically distributed points around the chief. Equation \eqref{eq:fov2} determines whether the point falls within the FOV of the sensor. The visual representation is depicted in Figure~\ref{fig:constraints}.

\subsection{{Illumination}}
The use of optical sensors for visual inspection requires a proper light source. That is, only illuminated points by lighting can be inspected through sensors.
In this work, the Sun is assumed to be the sole illumination source. To concentrate on the deputy's motion in the Hill frame, the Sun's motion is assumed to lie in the $\hat{\bf i}_c$-$\hat{\bf j}_c$  plane of the Hill frame. Thus, the Sun's position vector in the Hill frame is expressed by \cite{Wijk2024}
\begin{equation}\label{eq:sun}
    {\bf r}_s = d_{E-S}[\cos(\theta_s), \ \sin(\theta_s),\ 0]^T,
\end{equation}
where $d_{E-S}$ is the distance between the Sun and Earth, $\dot{\theta}_s=-n$, and $\theta_S$ is the angle between the Sun and the $\hat{\bf i}_h$ axis in the Hill frame. 
Considering the Sun's location, a point ${\bf p}_c$ is considered illuminated when the following condition{ is satisfied}:
\begin{equation}
    -1 < \frac{{\bf p_c}\cdot ({\bf p}_c-{\bf r}_s)}{||{\bf p_c}|| \cdot ||({\bf p}_c-{\bf r}_s)||} < 0.
\end{equation}
Note that only points within the tangential lines from the Sun are illuminated.


\section{Fuzzy-Driven Control and Optimization }\label{sec:control}
\subsection{Genetic Fuzzy System}
To {successfully} determine the {deputy's} force for inspection, this work {employs} the FIS, {an} AI approach {chosen for its} interpretability, design flexibility, and {compatibility} with optimization approaches. The FIS 
{operates in }three steps: fuzzification, inference, and defuzzification. 
Fuzzification {converts} numerical inputs into linguistic variables {(e.g., Negative, Zero, Positive)} using membership functions (MFs). The inference engine {then uses these linguistic input variables to} 
determines the output linguistic variables. 
{This step is highly transparent because} the rules that map 
the input {to} 
output 
are {expressed as clear }
If-Then statements. 
Finally, {defuzzification converts these linguistic outputs back into }output numeric values. 
{Traditionally}, the MFs {and} 
the rules 
are {set based on} 
expert knowledge of the system. 
However, this method does not guarantee an optimal solution. The complexity of designing these FIS components grows significantly with an increasing number of inputs and MFs, especially when the relationship between inputs and outputs is not straightforward. 
{To address this, }
this work incorporates {a genetic algorithm (GA), an optimization algorithm, to enable the FIS to learn and automatically determine its internal parameters. }
This {combined} framework is {known}
as a genetic fuzzy system. 
The GA {leverages}
its {robust} 
search capability within a {defined} 
search space to optimize the FIS parameters {without altering the fundamental FIS structure, making it a good choice for this task}. 

\subsection{Fuzzy-Driven Controller Design}
\begin{figure}[!b]
    \centering
    \begin{subfigure}[b]{0.49\textwidth}
        \centering
        \includegraphics[width=\textwidth]{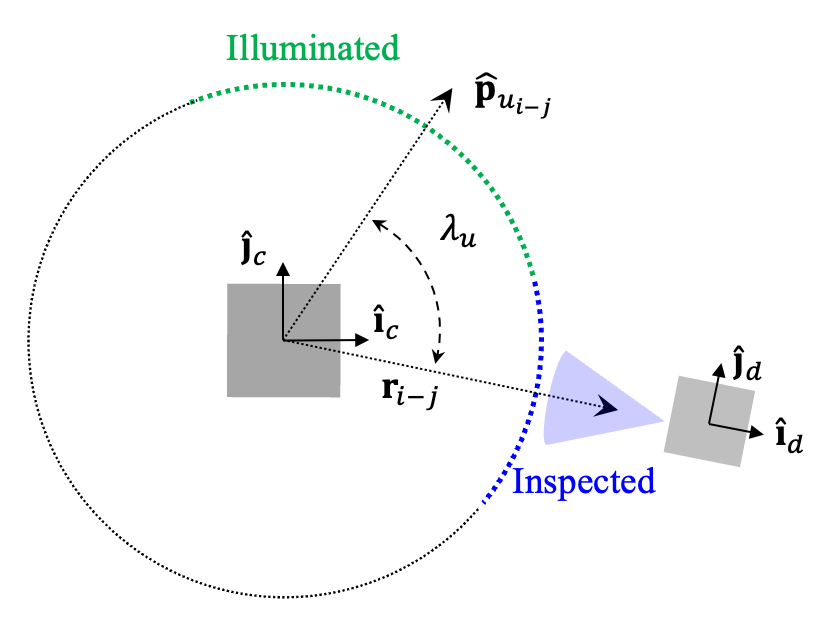}
        \caption{Projection on $\hat{\bf i}_c$-$\hat{\bf j}_c$ Plane}
        \label{fig:projection_xy}
    \end{subfigure}
    \begin{subfigure}[b]{0.49\textwidth}
        \centering
        \includegraphics[width=\textwidth]{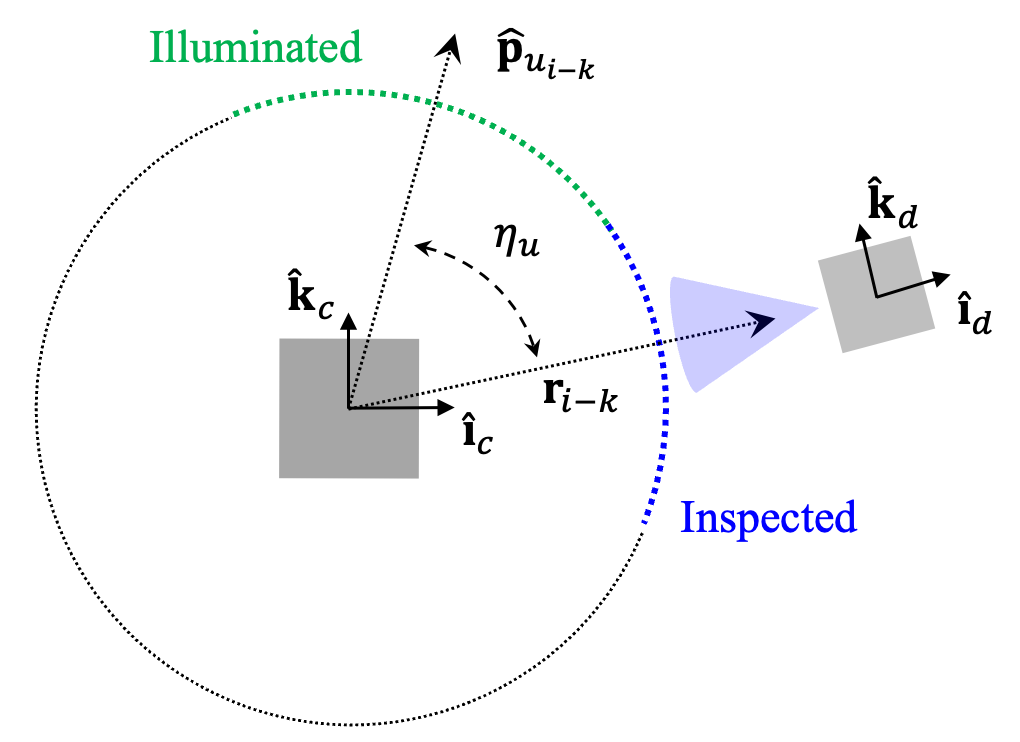}
        \caption{Projection on $\hat{\bf i}_c$-$\hat{\bf k}_c$ Plane}
        \label{fig:projection_xz}
    \end{subfigure}
    \caption{Projection on the Planes in the Hill Frame}
    \label{fig:projection}
\end{figure}

This work employs three FISs, each responsible for determining the deputy's force along a specific axis within the deputy's body-fixed frame. Since the relative translational motion is described in the Hill frame, the force vector (${\bf f}_d=[f_{dx},\ f_{dy},\ f_{dz}]^T\in\mathbb{R}^3$) {generated} in the deputy frame is then transformed into the Hill frame as ${\bf f} =[f_{x},\ f_{y},\ f_{z}]^T= C_{DC}^T{\bf f}_d.$

Each FIS takes two or three inputs, which include the relative distance, relative velocity, and relative angles, to determine the corresponding axial force.
Notably, the relative position and velocity information primarily influence the magnitude of the force, whereas the relative angular information dictates the deputy's direction movement.

As previously mentioned, the deputy can inspect the points that are both illuminated and within the optical sensor's FOV. Due to the FOV restriction, some illuminated points may remain uninspected at a given time step. For efficient inspection, it is advantageous to prioritize moving toward these uninspected but illuminated points. Therefore, a clustering technique is used to identify the centroid location of these uninspected and illuminated points, providing a directional vector defined as ${\bf p}_u$.

To obtain the necessary relative angles for determining each axis force, the vector ${\bf p}_u$ is projected onto two plains: the $\hat{\bf i}_c$-$\hat{\bf j}_c$ plane and $\hat{\bf i}_c$-$\hat{\bf k}_c$ plane. This yields two projection vectors, $\hat{\bf p}_{u_{i-j}}$ and $\hat{\bf p}_{u_{i-k}}$, respectively. Similarly, the projection vector of the deputy's position vector (${\bf r}$) onto each of these planes is also computed, as illustrated in Figure~\ref{fig:projection}. Consequently, using the two projection vectors, ${\bf r}_{i-j}$ and $\hat{\bf p}_{u_{i-j}}$, the relative angle $\lambda_{u}$ on the $\hat{\bf i}_c$-$\hat{\bf j}_c$ plane is computed, spanning a range from $-\pi$ to $\pi$. Concurrently, the relative angle $\eta_u$ on the $\hat{\bf i}_c$-$\hat{\bf k}_c$ plane is calculated within the same range. Hence, the state information and these derived relative angles are supplied as the input variables to the three FISs for determining each axis force in the deputy frame.

In the fuzzification process for each FIS, this work incorporates five MFs with associated linguistic variables for both each inputs and outputs. This ensures a sufficient level of granularity for the input data. To minimize the number of parameters within the MFs, Gaussian-shaped MFs are employed. The combination of each variable in the input can be represented as a table, where the table values correspond to the output linguistic variables. This table, in essence, forms the rule matrix, illustrating the relationship between the input and output variables. In this work, the means and standard deviations of certain MFs in the input and output, along with the rule matrices for all three FISs, are designed as tuning parameters to be optimized by the GA. 
These tuning parameters are consolidated into a single vector and then fed into the GA. Through the processes of selection, crossover, and mutation, the GA identifies the combination of parameters that yields the best-fit solution. 

To determine the optimal FIS parameters for the deputy's control force, the objective function introduced in Eq. \eqref{eq:obj} is re-defined for the GA by including the penalty term $\zeta$ as:
\begin{equation}
    \mathcal{L}_\text{fit} = \Delta v + \zeta.
\end{equation}
Note that $\zeta$ penalizes any undesirable behavior of the deputy. That is, the penalty is applied when the deputy violates the constraints associated with the collision avoidance and inspection rate.

\section{Simulation Study}
\subsection{Simulation Environment for Training and Testing}
\begin{table}[h!]
    \centering
    \caption{Simulation Parameters for the Chief and Deputy}
    \label{tab:sc}
    \begin{tabular}{c|c|c|c} \hline
        \multicolumn{3}{c|}{\textbf{Parameters}} & \textbf{Values (unit)} \\ \hline
        \multicolumn{2}{c|}{Simulation time} & $t_f$ & 3600 (s) \\
        \multicolumn{2}{c|}{Time interval} & $T$ & 10 (s) \\ \hline
         \multirow{6}{*}[1.2em]{Chief} 
         & Initial quaternion & ${\bf q}_c$ & $[0,\ 0,\ 0,\ 1]^T$ \\
         & Initial angular velocity & ${\bm\omega}_c$ & $[0,\ 0,\ n]^T$ {(rad/s)}\\ 
         & Distance {from} the chief {to} inspectable points & $d_s$ & 10 (m) \\
         & Mean motion & $n$ & 0.0011068 (rad/s) \\
         \hline
          \multirow{10}{*}{Deputy} 
         & Inertia & $J_d$ & $I_{3 \times 3}$ (kg$\cdot$ m$^2$) \\
         & Mass & $m_d$ & 12 (kg) \\
         & Initial relative quaternion & ${\bf q}$ & $[0,\ 0,\ 0,\ 1]^T$ 
         \\
         & Initial relative angular velocity & ${\bm\omega}$ & $[0,\ 0,\ n]^T$ (rad/s) \\ 
         & Inspection threshold & $\eta_\text{threshold}$ & 95 ($\%$) \\
         & Minimum relative distance & $d_\text{min}$ & 15 (m) \\
         & Maximum relative distance & $d_\text{max}$ & 200 (m) \\
         & Maximum force & ${f}_\text{max}$ & 1 (N) \\
         & {Maximum torque} & {$\tau_\text{max}$} & {10 (mNm)} \\
         & Half angle of field of view & $\beta$ & 15 (deg) \\
         & Boresight vector & ${\hat{\bf b}}$ & $[-1,\ 0,\ 0]^T$ \\
         \hline
    \end{tabular}
\end{table}
\begin{table}[h!]
    \centering
    \caption{Simulation Parameters for the GA}
    \label{tab:ga}
    \begin{tabular}{c|c}
    \hline
        \textbf{Parameters} &  \textbf{values} \\ \hline
        Number of population & 200 \\
        Maximum generation & 500 \\
        Selection function & Tournament selection \\ 
        Tournament size & 4 \\
        Crossover function & Scattered crossover \\ 
        Crossover rate & 0.8 \\
        Mutation function & Adaptive feasible mutation \\
        Mutation rate & 0.1 \\
        Elitism rate & 0.1 \\
        \hline
    \end{tabular}
\end{table}

The efficacy of the proposed fuzzy-driven control approach is verified through numerical simulations. For effective training (i.e., optimization) of the FISs responsible for determining the deputy spacecraft's control force, it is crucial to expose them to diverse environmental conditions. This ensures that the FISs can learn to adapt to various scenarios. As previously discussed, the FISs use relative information from the deputy as inputs. Consequently, to provide a variety of training situations, eight distinct initial relative position vectors for the deputy with respect to the chief are considered. These positions correspond to different sections within the Hill frame's planes, allowing the deputy to start its inspection from various starting points. For all training scenarios, the initial relative velocity is set to zero. Tables \ref{tab:sc} and \ref{tab:ga} detail the remaining simulation parameters for the chief and deputy spacecraft, as well as the optimization parameters for the GA. It is important to note that the GA training parameters are used for a single training iteration. Following the acquisition of newly trained FISs that demonstrate improved performance over their predecessors, an iterative training process is initiated. This process continuously refines the FIS parameters to achieve progressively lower fitness values.

For the testing phase, this study employs Monte Carlo simulations comprising 1,000 random initial conditions. To ensure a broad distribution of the deputy's initial relative positions around the chief and to simulate realistic inspection initiation, these random positions are selected such that the relative distance from the chief's center ranges from 50 to 100 m. This means the random initial positions are located within the region between two concentric spheres, with radii of 50 m and 100 m, respectively.

\subsection{Training Results}
\begin{figure}[!b]
    \centering
    \begin{subfigure}[b]{0.49\textwidth}
        \centering
        \includegraphics[width=\textwidth]{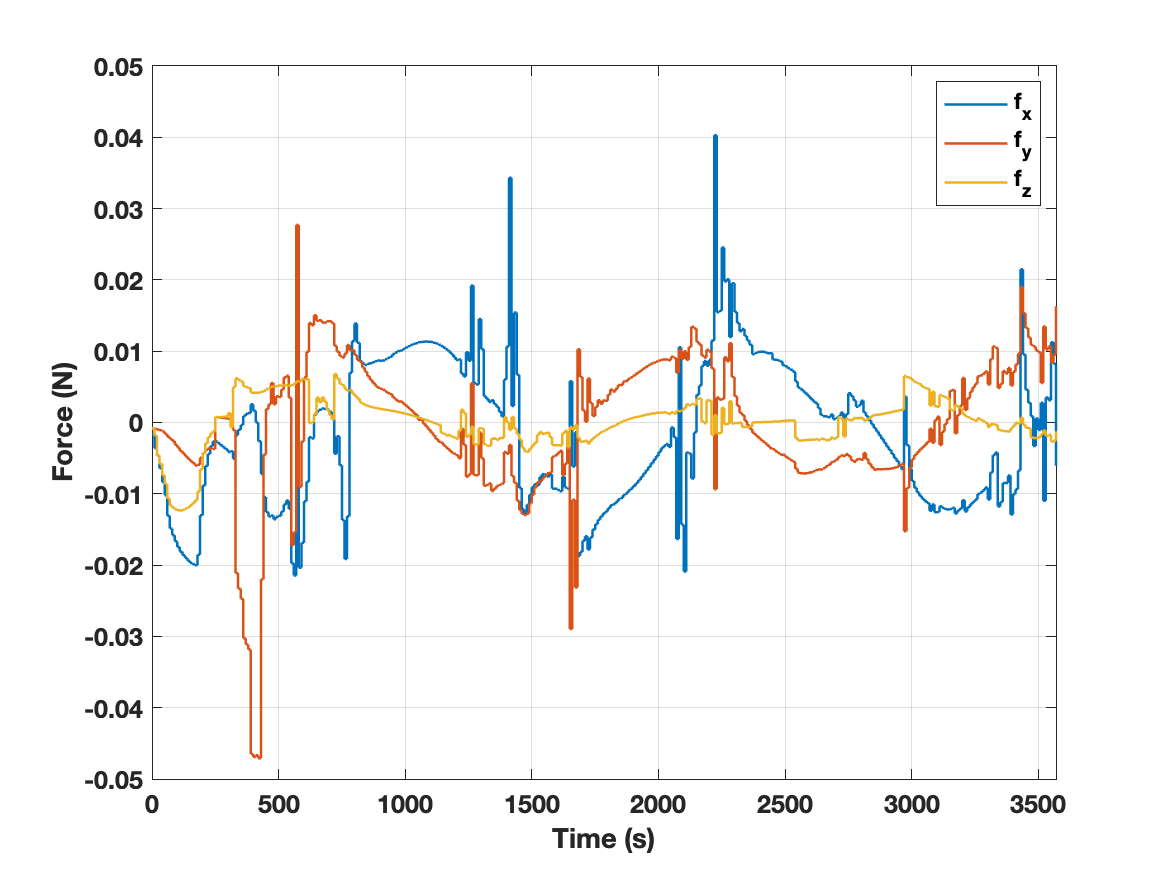}
        \caption{Control Force}
        \label{fig:train_sc1_force}
    \end{subfigure}
    \hfill
    \begin{subfigure}[b]{0.49\textwidth}
        \centering
        \includegraphics[width=\textwidth]{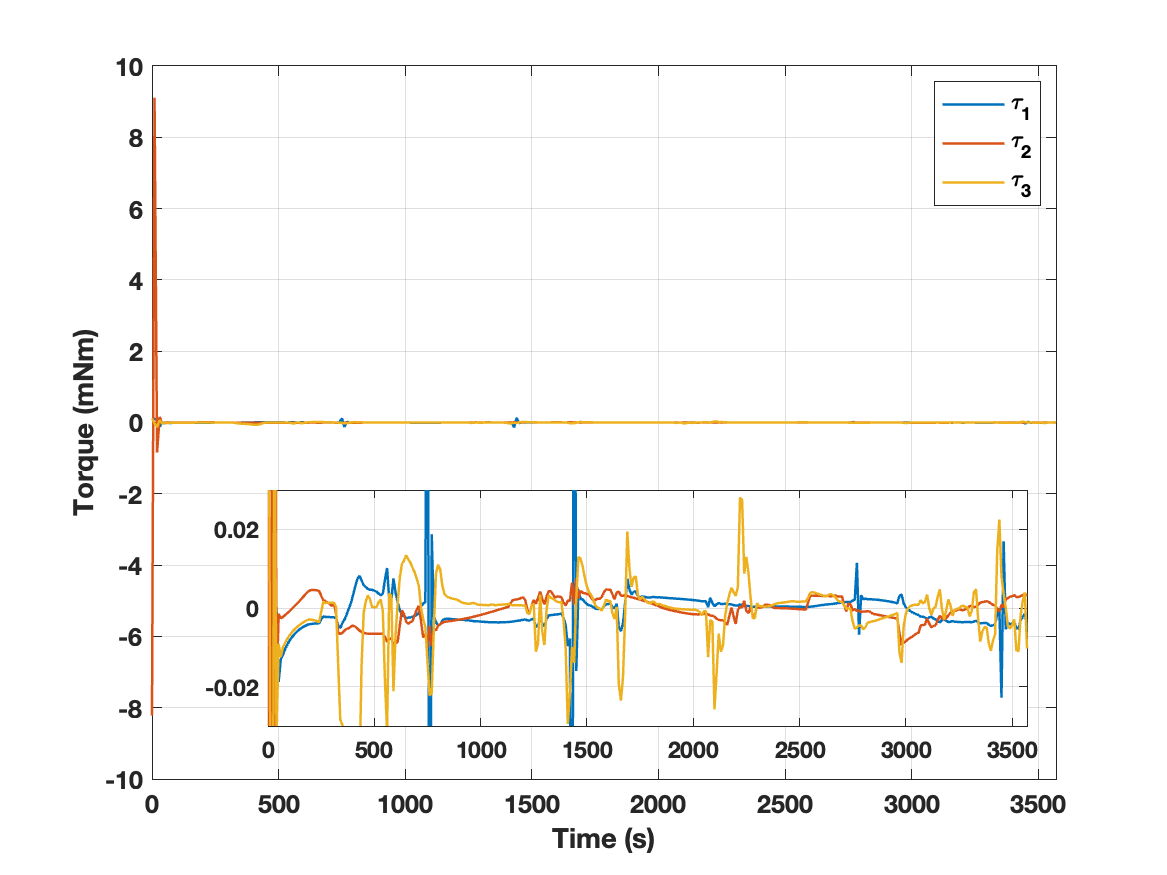}
        \caption{Control Torque}
        \label{fig:train_sc1_torque}
    \end{subfigure}
    \caption{{Deputy's Control Input}}
    \label{fig:train_sc1}
\end{figure}

After the iterative training, the optimized FISs are employed to generate the deputy's control force. Figure~\ref{fig:train_sc1} illustrates the deputy's control input information for a representative training scenario. The generated thrust force, maintained within $\pm$ 0.05 N, is well below the 1 N limit. It is notable the thrust force remains constant between each time step. 

\begin{figure}[!b]
    \centering
    \includegraphics[width=0.7\linewidth]{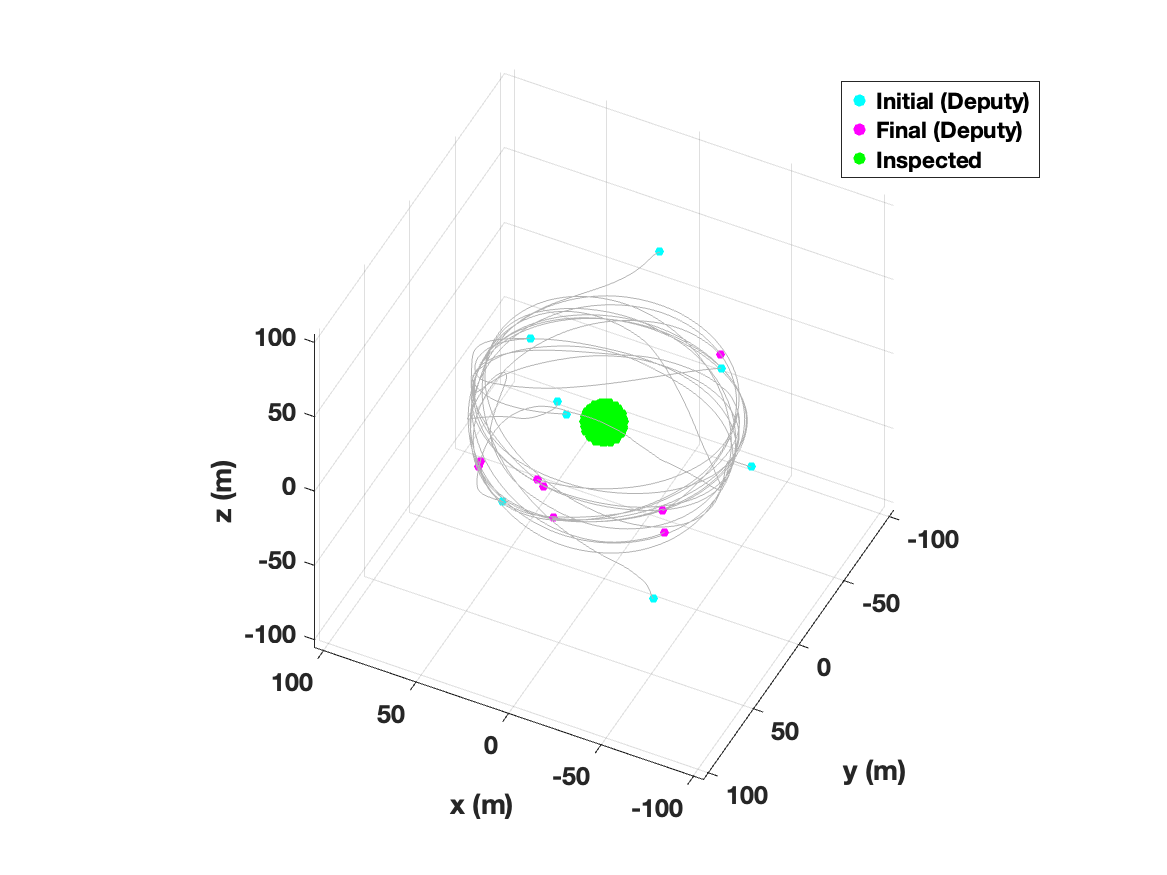}
    \caption{Deputy's Inspection Trajectory}
    \label{fig:train_traj}
\end{figure}

For the attitude control torque, most torque consumption occurs at the maneuver's initiation, as the deputy performs attitude alignment to point towards the chief. Once the deputy is aligned with the chief, only minimal control torque is subsequently required. The remaining training scenarios have similar trend without actuation limit violations.

Figure~\ref{fig:train_traj} displays the overlapped 3-dimensional (3D) trajectories, showing circular-like patterns. 
This indicates that the FISs are effectively trained to balance the inspection rate and fuel consumption, enabling the deputy to maintain a nearly circular relative orbit for efficient inspection.

\begin{figure}[t!]
    \centering
    \includegraphics[width=1\linewidth]{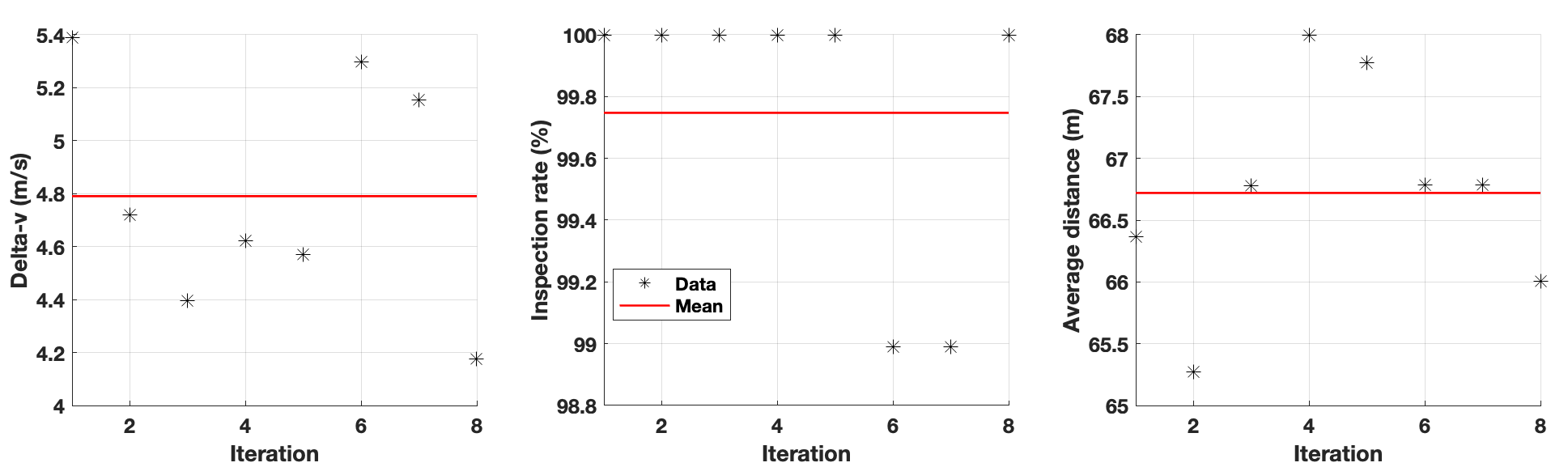}
    \caption{Training Results Distributions}
    \label{fig:train_dist}
\end{figure}
\begin{table}[t!]
    \centering
    \caption{Summary of the Training Results}
    \label{tab:train}
    \begin{tabular}{c|c|c|c}
        \hline
        \multirow{2}{*}{8 Training scenarios} & $\Delta v$ consumption & Inspection rate & Average relative\\ 
         & (m/s) & (\%) & distance (m) \\ \hline
        Mean & 4.799 & 99.748 & 67.030 \\
        Standard deviation & 0.458 & 0.468 & 1.280 \\ \hline
    \end{tabular}
\end{table}

Furthermore, the performance of the trained FISs is evaluated across all training scenarios in terms of $\Delta v$ consumption, inspection rate, and average relative distance, as shown in Figure~\ref{fig:train_dist} and summarized in Table~\ref{tab:train}. 

The results demonstrate that the deputy of 12 kg (like a CubeSat sized between 6U and 12U with a limited FOV) successfully inspects the chief, which has a diameter of 20 m, while consuming an average of about 4.799 m/s of $\Delta v$ and maintaining an average relative distance of approximately 67.03 m from the chief. Crucially, the deputy nearly completes the inspection within the one-hour timeframe, consistently exceeding the 95 \% inspection threshold.
This demonstrates that the FISs are well-trained and operate within all specified constraints.

\subsection{Testing Results}
\begin{figure}[t!]
    \centering
    \includegraphics[width=0.8\linewidth]{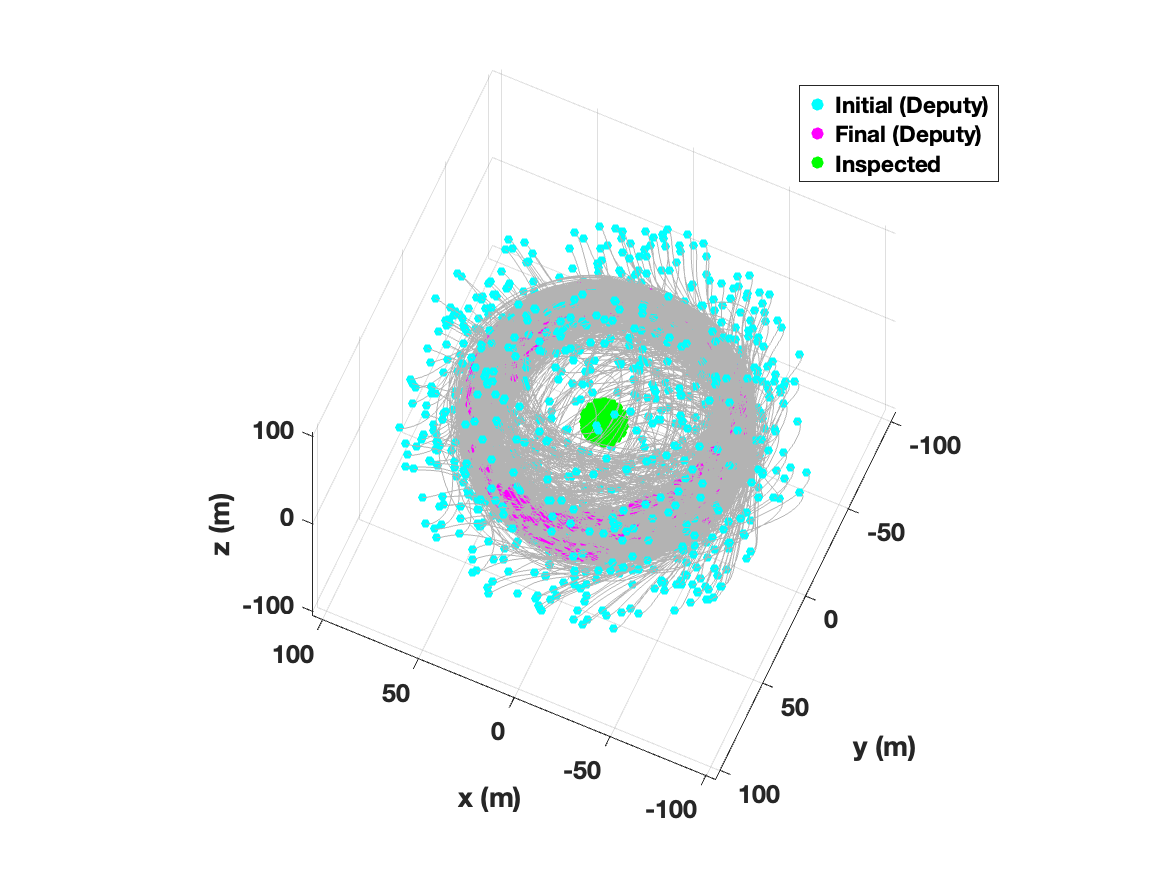}
    \caption{Deputy's Inspection Trajectory Overlapped (1,000 cases)}
    \label{fig:test_traj}
\end{figure}

In the testing phase, the trained FIS-driven controller's performance is validated by deploying the deputy from 1,000 random initial relative positions. 
Figure~\ref{fig:test_traj} presents the overlapped trajectories of the deputy starting from these diverse initial conditions. 
Similar to the training results, circular-like patterns around the chief are prominently observed. 

\begin{figure}[t!]
    \centering
    \includegraphics[width=1\linewidth]{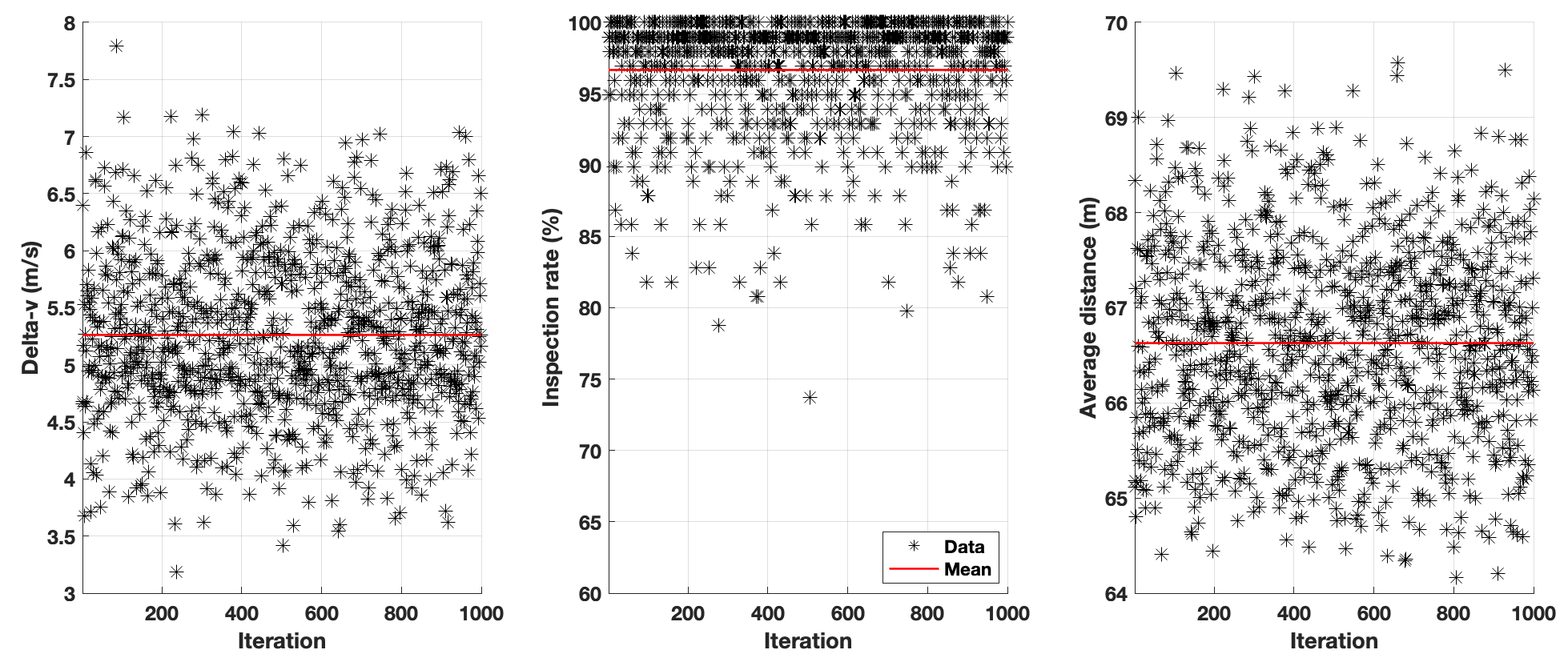}
    \caption{Testing Results Distribution}
    \label{fig:test_dist}
\end{figure}

Additionally, the evaluation results for $\Delta v$ consumption, inspection rate, and average relative distance from the testing are tabulated in Table \ref{tab:test} and visually represented in Figure \ref{fig:test_dist}. Figure \ref{fig:test_dist} specifically depicts the distribution of these testing results, showing that most data points clusters around the mean, despite the presence of some outliers. 
While the $\Delta v$ consumption is slightly higher than in the training, the mean inspection rate across all 1,000 random cases remains high, exceeding the 95 \% inspection threshold. In addition, the average relative distance of the deputy from the chief closely matches the values observed during training. 

\begin{table}[t!]
    \centering
    \caption{Summary of the Testing Results}
    \label{tab:test}
    \begin{tabular}{c|c|c|c}
        \hline
        \multirow{2}{*}{1,000 Testing scenarios} & $\Delta v$ consumption & Inspection rate & Average relative\\ 
         & (m/s) & (\%) & distance (m) \\ \hline
        Mean & 5.62 & 96.667 & 66.629 \\
        Standard deviation & 0.715 & 3.919 & 1.039 \\ \hline
    \end{tabular}
\end{table}


\section{Conclusions}
This study presents the design of a fuzzy inference system (FIS)-driven control approach aimed at achieving minimal fuel consumption during autonomous spacecraft inspection. The approach considers various constraints, including actuator limitations, collision avoidance, reliable inspection requirements, restricted sensor field of view, and illumination conditions.
By integrating a bio-inspired optimization technique, the fuzzy-based controller is optimized to minimize fuel usage while strictly adhering to these constraints. The resulting FIS determines the control force, which generates a circular-like inspection trajectory around the chief spacecraft. This trajectory ensures 
acceptable inspection while preventing collisions. Extensive Monte Carlo simulations demonstrate that the trained controller achieves an inspection success rate exceeding 96\%, with only a slight increase in fuel consumption compared to the training results.
For future work, it is planned to incorporate nonlinear relative motion and account for environmental disturbances. This includes considering factors, such as light reflection and optical sensor's keep-out constraints, to enhance the practicality of the approach.

\bibliographystyle{AAS_publication}   
\bibliography{references}   

\end{document}